\documentclass[fleqn,12pt]{wlscirep}
\usepackage[utf8]{inputenc}
\usepackage[T1]{fontenc}
\usepackage{lineno}
\usepackage{float}
\title{Hybrid Metasurfaces for Simultaneous Focusing and Filtering}

\author[1,]{Mansoor A. Sultan}
\author[2]{Fatih Balli}
\author[1]{Daniel L. Lau}
\author[1,*]{J. T. Hastings}
\affil[1]{Department of Electrical and Computer Engineering, University of Kentucky, Lexington, Kentucky 40506, USA}
\affil[2]{Department of Physics and Astronomy, University of Kentucky, Lexington, Kentucky  40506, USA}

\affil[*]{Corresponding author: todd.hastings@uky.edu}

\begin{abstract}
This work presents the design and fabrication of simple, polymeric, structure-based optical filters that simultaneously focus light. These filters represent a novel design at the boundary between diffractive optics and metasurfaces that may provide significant advantages for both digital and hyperspectral imaging. The fabrication process for the proposed filters resembles 3D printing, and is based on direct laser writing of a polymeric material using two-photon lithography. In addition, printed structures could be used to create molds for nanoimprint replication and mass production. Filters for visible and near-infrared wavelengths were designed using finite difference time domain (FDTD) simulations. 
\end{abstract}
\begin{document}

\flushbottom
\maketitle

\thispagestyle{empty}

Color filters are essential for a wide range of applications such as digital cameras, projectors, displays, and spectral imaging systems. However, most of these applications still utilize tri-color dye- or pigment-based filters. These filters transmit  the  desired spectral region and absorb the rest. The material composition of these filters make them vulnerable to environmental conditions, especially photo- and thermal- degradation\cite{inaba_degradation-free_2006}. In imaging systems, the integration of these filters with a sensor requires three steps of aligned lithography.  This leads to a protracted and expensive fabrication process. Therefore, structure-based color filters have been of great interest because of their durability, wide tunability, and simplicity of integration with imaging systems. 

Several types of structural filters have been investigated as a replacement for color pigments in printing and imaging. These include plasmonic \cite{ema_plasmonic_2018,hwang_plasmonic_2018}, Metal Insulator Metal (MIM) \cite{wang_stepwise-nanocavity-assisted_2018}, Guided-Mode-Resonance (GMR) \cite{hemmati_fiber-facet-integrated_2018}, Photonic Crystal (PhC) \cite{srinivasan_micro-optical_2009}, and color routing structures \cite{chen_gan_2017}. However, each of these filter designs have shortcomings. For instance, the plasmonic color filters suffer from ohmic losses associated with the metallic structure in their design, and GMR filters are very sensitive to angle of incidence. The MIM and PhC filters require multi-step fabrication processes with high fabrication cost related to precision alignment and gray-scale lithography. Moreover, the transmittance of both plasmonic and MIM filters are very low which consequently degrades the sensor sensitivity. Finally, color routing devices have high cross-talk between the color pixels.

Recently, tremendous effort has been invested in light modulation by thin film materials patterned in quasi-periodic structures called metasurfaces. In contrast to conventional optics, metasurfaces do not depend on the phase modulation by thick lenses to optimize the light path.  Instead, metasurfaces introduce subwavelength nanostructures that allow controllable phase shifting of the propagating light. Metasurfaces can offer precise control over light intensity, phase, and polarization; thus, they have been widely utilized in the design and fabrication of optical metafilters \cite{cheng_structural_2015,bruckner_inverted_2014}, metalenses \cite{khorasaninejad_visible_2017, ye_silicon-rich_2019,chung_high-na_2019,chen_high_2018,balli_hybrid_2020},and in vortex beam generation \cite{ma2017trapping,Kotlyar_2018,balli2020rotationally}.

Two-dimensional diffraction gratings composed of sub-wavelength pillars have also been utilized in structural color printing and filtering \cite{nawrot_transmission_2013,catchpole_conceptual_2007}. These structures lie at the boundary between diffractive optics and metasurfaces.  They exploit waveguiding effects in the pillars to control the coupling of incident light to the various diffracted orders, including the transmitted and reflected zero orders.  As a result, they can produce tailored transmission spectra, especially when the transmitted light is angularly filtered by a subsequent lens or spatially filtered by an aperture.  

In this work, we present a novel approach to creating all-dielectric color filters that also focus light.  These hybrid filters exploit a combination of diffraction by periodic structures, phase shifting by a phase plate, and waveguiding by subwavelength structures to simultaneously focus and filter light in the visible region of the spectrum. These filters can be easily fabricated with two-photon lithography (TPL), a technique that has been widely used in the fabrication of complex, sub-micron, 3D structures. The flexibility of fabrication by TPL has been proven in different applications with feature size approaching the diffraction limit \cite{kawata2001finer} such as photonic crystals\cite{liu_structural_2019},microlenses \cite{thiele_3d-printed_2017}, and waveguides \cite{durisova_ip-dip_2018}. The reported filters have the further advantage that they have no reentrant features and could be replicated by molding for high volume production.

\begin{figure}[H]
\begin{center}
 \includegraphics[width=0.8\textwidth]{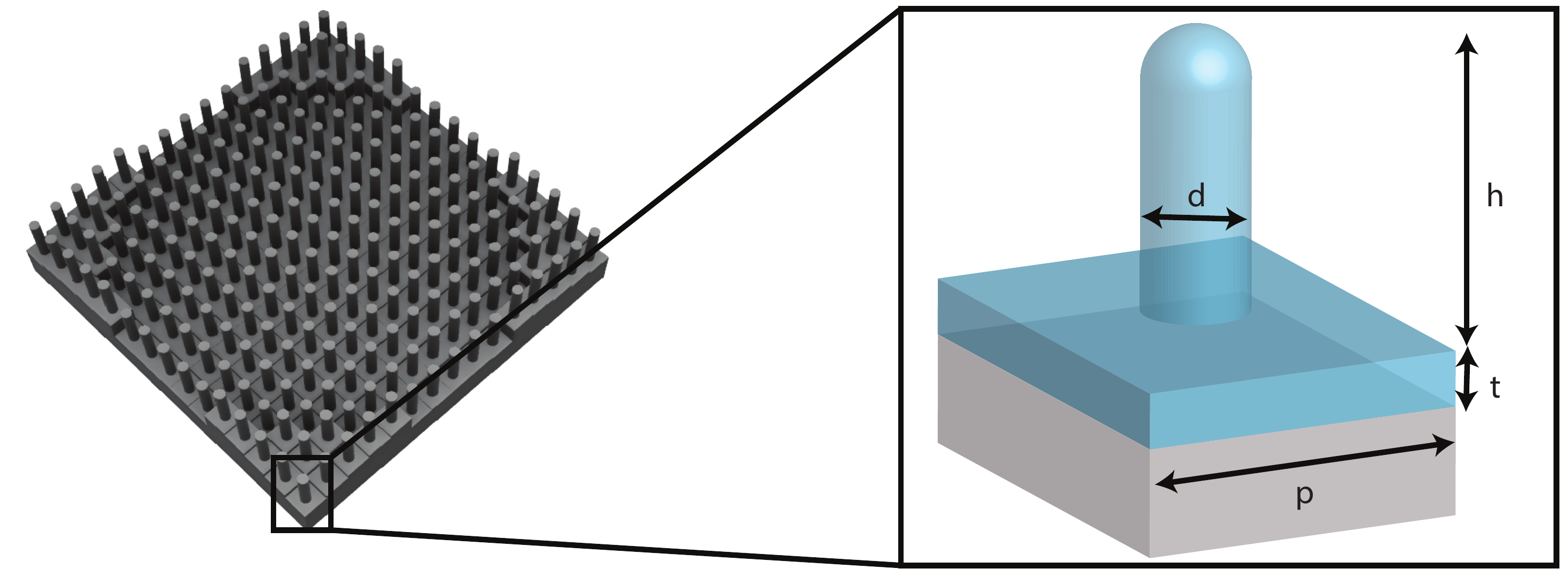} 
    \caption{3D model of a hybrid metasurface filter and its central unit cell with the variable dimensions labeled.}
    \label{fig1}
    \end{center}
\end{figure}

Here, we implement hybrid metasurface filters based on the structure shown in Fig. \ref{fig1}. The central unit consists of a phase-plate and pillar with four degrees of freedom: periodicity, phase-plate thickness, pillar diameter, and pillar height. Lim \textit{et al.} and Balli \textit{et al.} utilized a similar concept for holographic color prints and hybrid metalens designs, respectively \cite{lim_holographic_2019,balli_hybrid_2020}.  Lim \textit{et al.} used a fixed phase-plate thickness and changed the pillar height for each holographic color pixel.  In contrast, we vary the phase-plate thickness to obtain focusing behavior, while keeping the pillar height constant to control the filter response.  

In order to focus specific wavelengths at a focal length of $\boldsymbol{f}$, the transmitted light must have a parabolic phase profile that satisfies 
\begin{equation}
    \phi(r,\lambda)=-\frac{2\pi}{\lambda}(\sqrt{r^{2}+f^{2}}-f)
    \label{eq:1}
\end{equation}
where $r$ is the radial coordinate. The phase delay generated by light propagation in a film with thickness, $t$, and refractive index, $n$, is given by 
\begin{equation}
    \phi(\lambda)=\frac{2\pi (n-1)t }{\lambda}.
    \label{eq:2}
\end{equation}
Equation (\ref{eq:1}) represents the required spatial phase shift to focus the light, and Eq. (\ref{eq:2}) represents the spectral phase shift due to light propagation in a film.  From Eqs. (\ref{eq:1}) and (\ref{eq:2}), the required phase-plate thickness can be obtained from
\begin{equation}
    t(r,\lambda)=\frac{\bmod(\phi,2\pi)}{\frac{2\pi}{\lambda}(n-1)}.
    \label{eq:3}
\end{equation}
The resulting phase-plate will be a quantized Fresnel lens structure. 

\begin{figure}[H]
\begin{center}

    \includegraphics[width=0.6\textwidth]{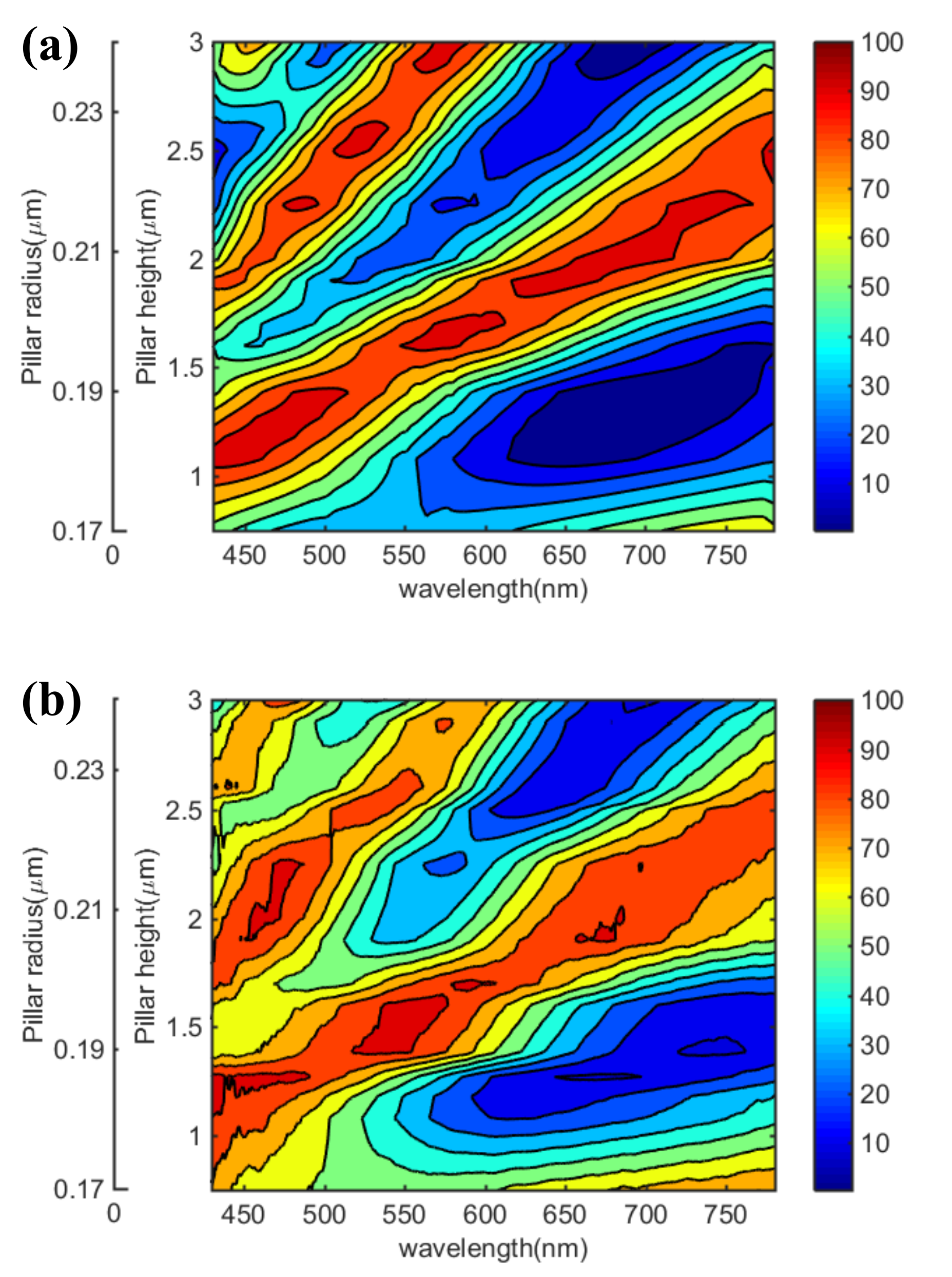} 
    \caption{Transmittance of pillars arrays (non-focussing) as a function of wavelength, pillar radius, and pillar height. \textbf{(a)} Simulated transmittance \textbf{(b)} Measured transmittance. The periodicity was fixed at 1.1 $\mu$m.}
    \label{fig2}
    \end{center}
\end{figure}

The 2D grating in combination with the phase-plate works as filter. The filtering process is based on diffraction and waveguiding effects that introduce phase differences between the light that propagates inside and outside the pillars\cite{lim_holographic_2019,nawrot_transmission_2013}. The design process was implemented using the finite-difference time-domain (FDTD) method. First, the pillars were simulated to estimate the dimensions that correspond to the desired filter function. A unit cell of the 2D grating on a fused silica substrate was simulated with periodic boundary conditions for $\boldsymbol{x}$ and $\boldsymbol{y}$ and perfectly matched layers on top and bottom. A plane wave light source was placed inside the substrate and $2 \mu$m below the pillar, and a frequency domain field and power monitor was placed $12 \mu$m above the pillar to collect transmitted power. For the experiments discussed below, only light with $<30^{\circ}$ diffraction angle will be collected by the objective. 
The results were calculated and plotted for different pillars height and radius as shown in Fig. \ref{fig2}(a). 

In order to test the simulated structure, pillar arrays were fabricated with different heights and laser powers to optimize the dimensions. Fig.\ref{fig2}(b) shows the transmittance of fabricated pillar arrays when the average laser power was set at 37.5 mW for the structure was sliced by  0.2 $\mu$m. The fabrication results agree well with the simulations, and the filter function can be predicted based on pillar height and radius. The radius is a function of laser power and pillar height and it is expected that the pillar radius increases with height as more voxels overlap in taller pillars.  For the wavelength range of interest, shorter and narrower pillars yield short-pass filters.  As the pillars become taller, the filter transitions to a band-pass response, then to a notch-like response, and finally to oscillatory response.

Second, we generated the required phase library for the phase-plate thickness to obtain the required phase profile as in Eq. (\ref{eq:2}). The periodicity of the pillars was fixed at 1.0 $\mu m$, and their height and radius were optimized for the desired filter function based on Fig. \ref{fig2}. The filters were designed on a fused silica substrate. IP-Dip resist, the two photon lithography resist from Nanoscribe GmbH, was used in simulation process for phase-plate and pillars structures. The fused silica refractive index was obtained from \cite{malitson_interspecimen_1965} and the exposed IP-Dip resist refractive index was from the Nanoscribe material information database \cite{Ipdip}. The final filter design has a focal length of 43 $\mu m$ and entrance pupil diameter of 7 $\mu m$. The pillar heights were 0.9, 1.4 and 1.8 $\mu m$ for the three desired colors: blue, green and red respectively.

Two-photon lithography was employed to fabricated the hybird metasurface filters. Two photon lithography is a 3D lithography process that generates complex structure in a single step. The process utilizes the absorption of two photons with low energy that cross-link the resist in the focal volume of the laser beam. The TPL system is a Photonic Professional GT (Nanoscribe GmbH) that uses a femtosecond laser with 780 nm wavelength. The fabrication process was conducted in dip-in mode with 63x objective (1.4 NA). A laser power sweep was performed for both the pillars and the phase-plate to optimize for the best filters performance. Scanning electron microscopy (SEM) images of the fabricated hybrid metasurface filters are shown in Fig. \ref{fig3}(a-c).     

Two light sources were used in the test setup (shown in Fig. S1 in supplementary material): a broadband light source (tungsten-halogen lamp) for focusing efficiency and pillar transmittance measurements over the visible range and a super-continuum laser source (SuperK Extreme with SuperK Select, NKT Photonics) to measure the full-width at half maximum (FWHM) at center wavelengths of 450~nm, 550~nm, and 650~nm. A collimating lens was add during FWHM measurement and removed for pillar characterization and focusing efficiency measurement. Focusing efficiency and pillar transmittance were measured by a fiber-coupled spectrophotometer (Ocean Optics HR4000CG-UV-NIR). The collection area of the fiber was adjusted to ensure that light was collected only from the focused spot.  The FWHM was measured by a high resolution Ximea grey scale camera and the image were post-processed with MATLAB.  

The wavelength dependent focusing efficiency matched well with the design as shown in Fig. \ref{fig3}. The fabricated filters have focusing efficiency of 50-55\% as shown in Fig. \ref{fig3}(d-f). In comparison to plasmonic color filters\cite{fleischman_hyper-selective_2017,yokogawa_plasmonic_2012}, these filters offer higher transmittance by virtue of the the dielectric material from which they are made.

The FWHM of the focused spots shows excellent agreement with simulation, as shown in Fig. \ref{fig4}(a-f).  The finite bandwidth of the source does not contribute significantly to the spot size.  Likewise, the FWHM varies with wavelength as expected.  The spot sizes themselves are well matched to common digital camera sensor pixel sizes regardless of wavelength.  Thus, with refinement, similar filters might ultimately be integrated with CCD or CMOS image sensors\cite{kato_320_2018}. 
As mentioned earlier, hybrid filters are based on both diffraction and waveguiding. However, the results presented in Fig. \ref{fig4} do not show the higher diffracted orders because these waves are beyond the numerical aperture of the objective (NA = 0.50). However,  higher order diffraction is visible in the simulated optical power distributions shown in Fig. \ref{fig5}.  These simulations confirm  that higher orders will not be collected with numerical aperture that we employed. (For more detail, see the far-field angular distributions in Fig. S2.)  Thus, unlike non-focusing filters of similar design, either spatial or angular separation of the diffracted light can be used with the focusing filters.  In addition, changing the angle of incidence reduces the integrated transmission of the filters, but does not dramatically change the shape of the filter spectrum as shown in the supplementary Figs. S3 and S4.  Thus, the filters maintain their color performance even off axis.

In conclusion, we have demonstrated the design and fabrication of novel structure-based color filters that also focus light. The filters have focusing efficiency of  50-55\%. These filters overcome some of the shortcomings of other filters such as plasmonic, GMR, PhC, MIM and metafilters, as long as higher order diffraction can be separated. They offer zero ohmic losses, a simpler fabrication process, and their spectral response does not strongly depend on incident angle.  The focused spot size is smaller than both the filter itself and common image sensor pixel sizes.  The focusing characteristic of these hybrid metasurface filters offers another appealing property that could make refined designs a potential replacement for the filter and microlens layers in modern camera sensors. \\

\noindent\textbf{{Funding.}} Intel Corporation; National Science Foundation (IIS-1539157).
\newline
\noindent\textbf{{Acknowledgement.}} The authors thank the UK Center for Nanoscale Science and Engineering, a member of the National Nanotechnology Coordinated Infrastructure (NNCI), which is supported by the National Science Foundation (ECCS-1542164). This work used equipment supported by National Science Foundation Grant No. CMMI-1125998.\\
\noindent\textbf{Disclosures.} The authors declare no conflicts of interest.
  \bibliography{main}
  \newpage
%\section{figures}

\begin{figure}[H]
\begin{center}
    \includegraphics[width=0.7\textwidth]{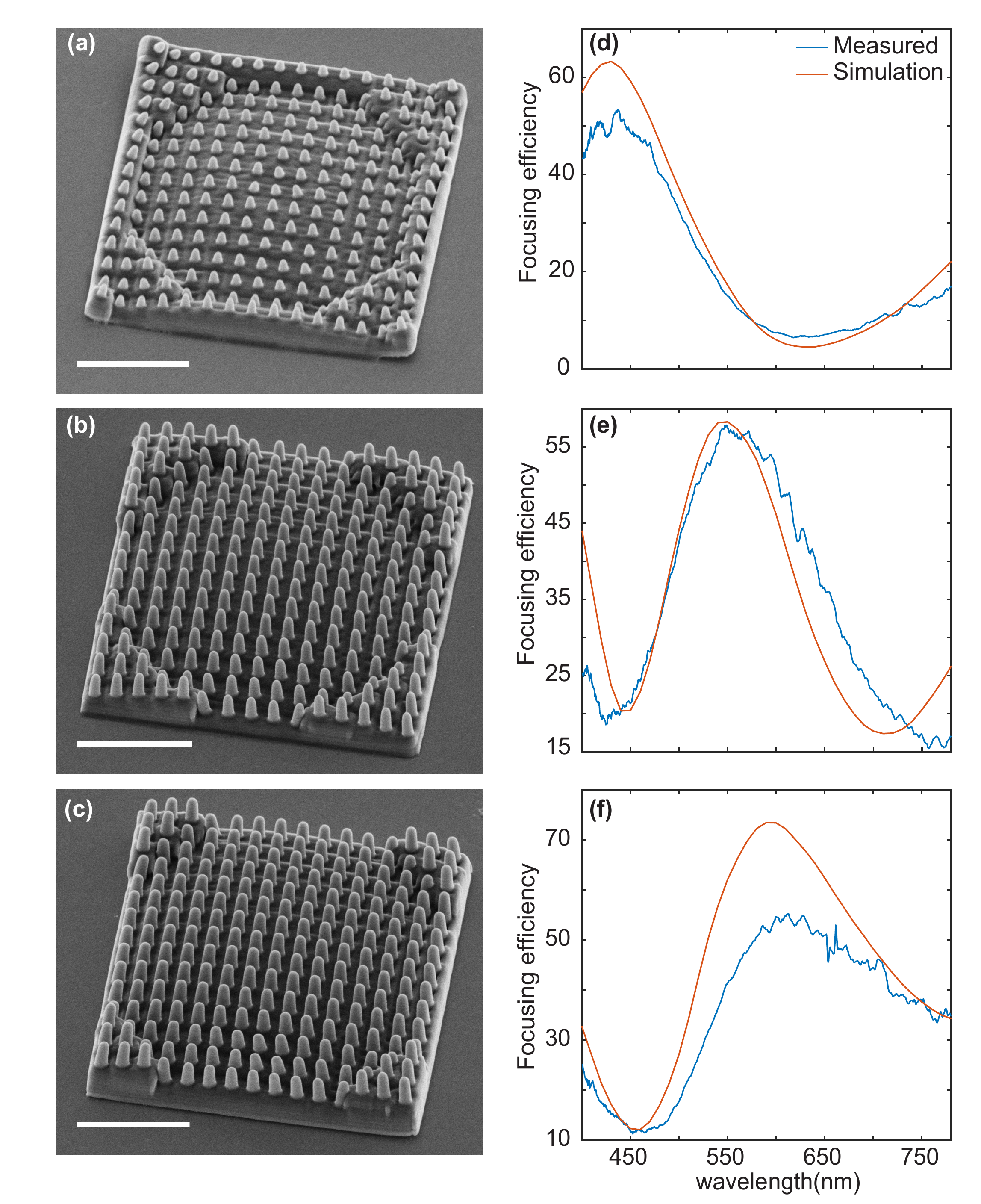}
    \caption{\textbf{(a-c)} SEM images for 3D printed hybrid metasurface filters with peaks in the blue, green and red respectively. The scale bar for all images is 5$ \mu m$ \textbf{(d-f)} Simulated (red) and measured (blue) transmittance for the fabricated filters over the visible spectrum for the three filters (blue, green and red respectively).}
    \label{fig3}
    \end{center}
\end{figure}

\begin{figure}[H]
    \begin{center}
     \includegraphics[width=0.8\textwidth]{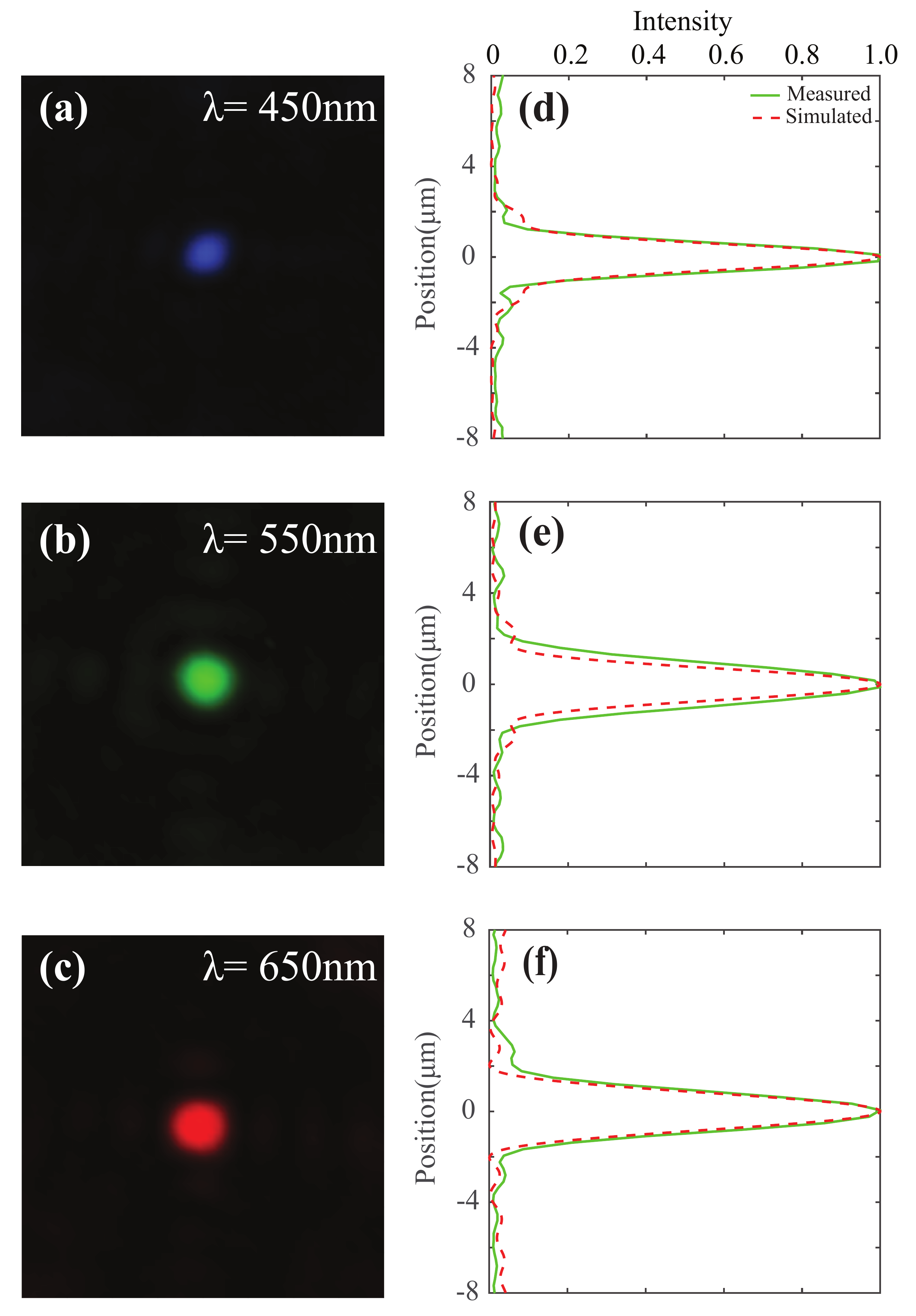} % will be changed later
    \caption{FWHM characterization for the fabricated filter.\textbf{(a-c)} Intensity distribution in the focal plane for Blue, Green, and Red respectively. \textbf{(d-f)} are comparison between the simulated and measured relative intensity in(a.u) at the focal plane for Blue, Green, and Red respectively.}
    \label{fig4}
        \end{center}
\end{figure}

\begin{figure}[H]
    \begin{center}
     \includegraphics[width=\textwidth]{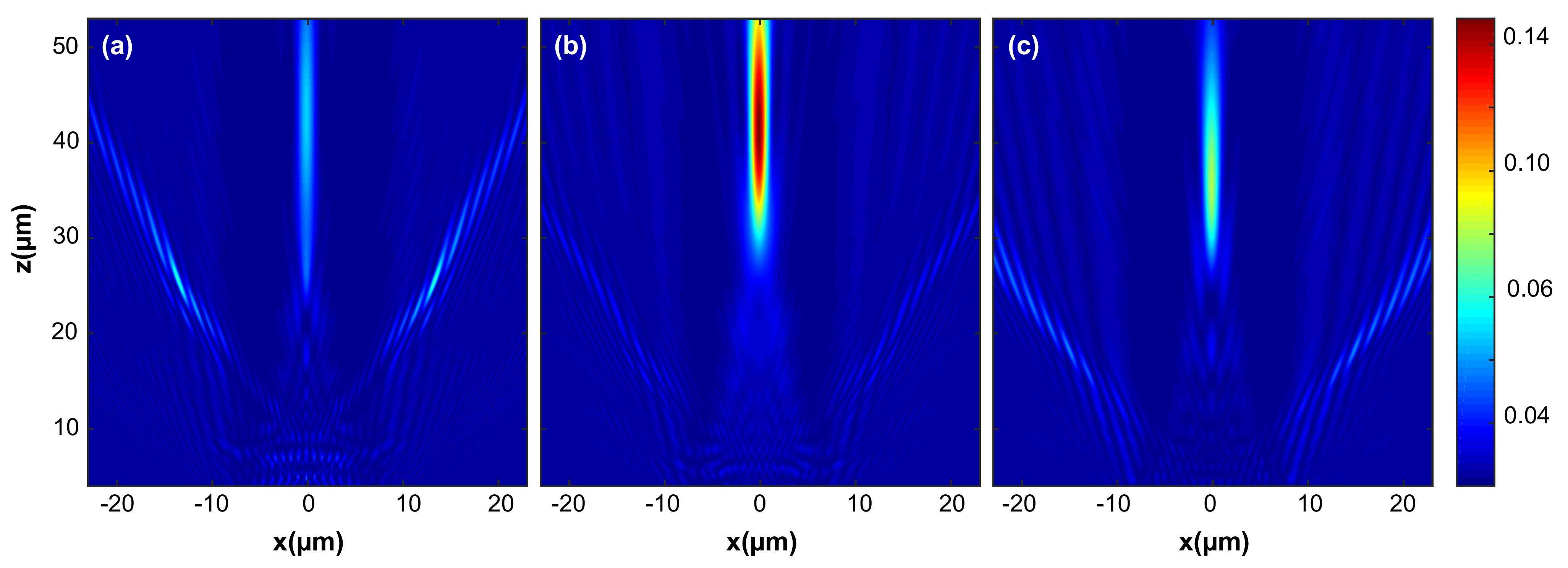}
    \caption{Simulated optical power distribution of hybrid filters showing the higher diffracted orders of the blue filter at RGB wavelengths \textbf{(a)} 450nm, \textbf{(b)} 550nm, \textbf{(c)} 650nm.}
    \label{fig5}
        \end{center}
\end{figure}
\newpage
\newcommand{\beginsupplement}{%
        \setcounter{table}{0}
        \renewcommand{\thetable}{S\arabic{table}}%
        \setcounter{figure}{0}
        \renewcommand{\thefigure}{S\arabic{figure}}%
     }
     
\section*{\LARGE{Supplementary Material : Hybrid Metasurfaces for Simultaneous Focusing and Filtering}}
\beginsupplement
\section{Experimental Setup}
\begin{figure}[H]
\begin{center}
        \includegraphics[width=0.7\textwidth]{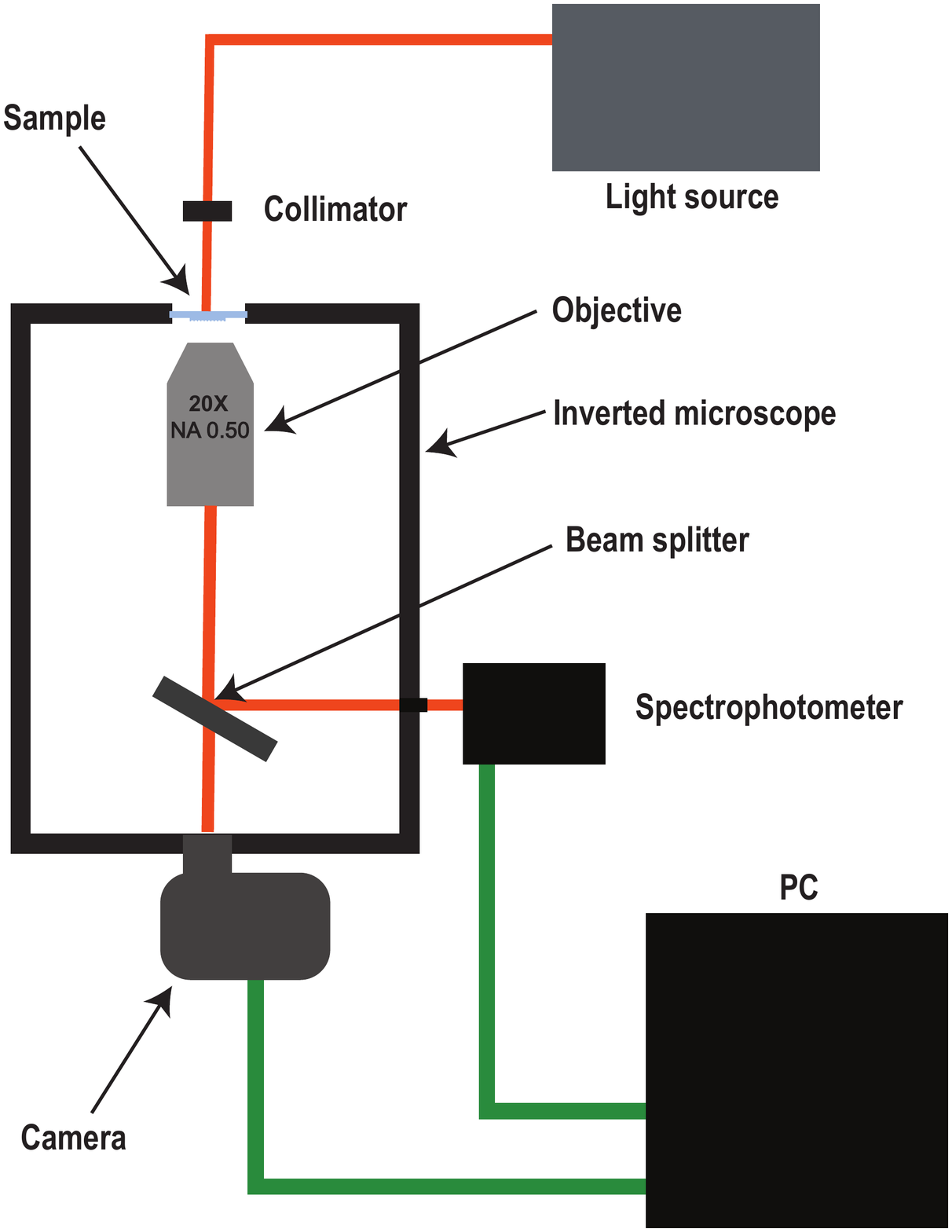}[H]
    \caption{Experimental setup for filter characterization}
   \label{figs1}
\end{center}
\end{figure}

The optical setup for filter characterization is shown in Fig. \ref{figs1}. The light source is halogen-deuterium fiber coupled source from Ocean Optics. Two types of camera were used: a colored camera (Amscope MU100) and gray-scale camera (Ximea - xiQ). The spectrophotometer is a UV-NIR model (HR4000CG-UV-NIR) from Ocean Optics.

\section{Fabrication Process}
The filters were fabricated on fused silica substrates of 0.7mm thickness with IP-DIP resist (Nanoscribe GmbH). After exposure the sample was developed for 20 min with 2-Methoxy-1-methylethyl acetate (PGMEA from J.T. Baker) then rinsed for 3 min in IPA and 30 seconds in Engineering fluid.  
\section{SUPPLEMENTARY SIMULATIONS}
\subsection{Higher diffracted orders}
\begin{figure}
\begin{center}
    \includegraphics[width=0.5\textwidth]{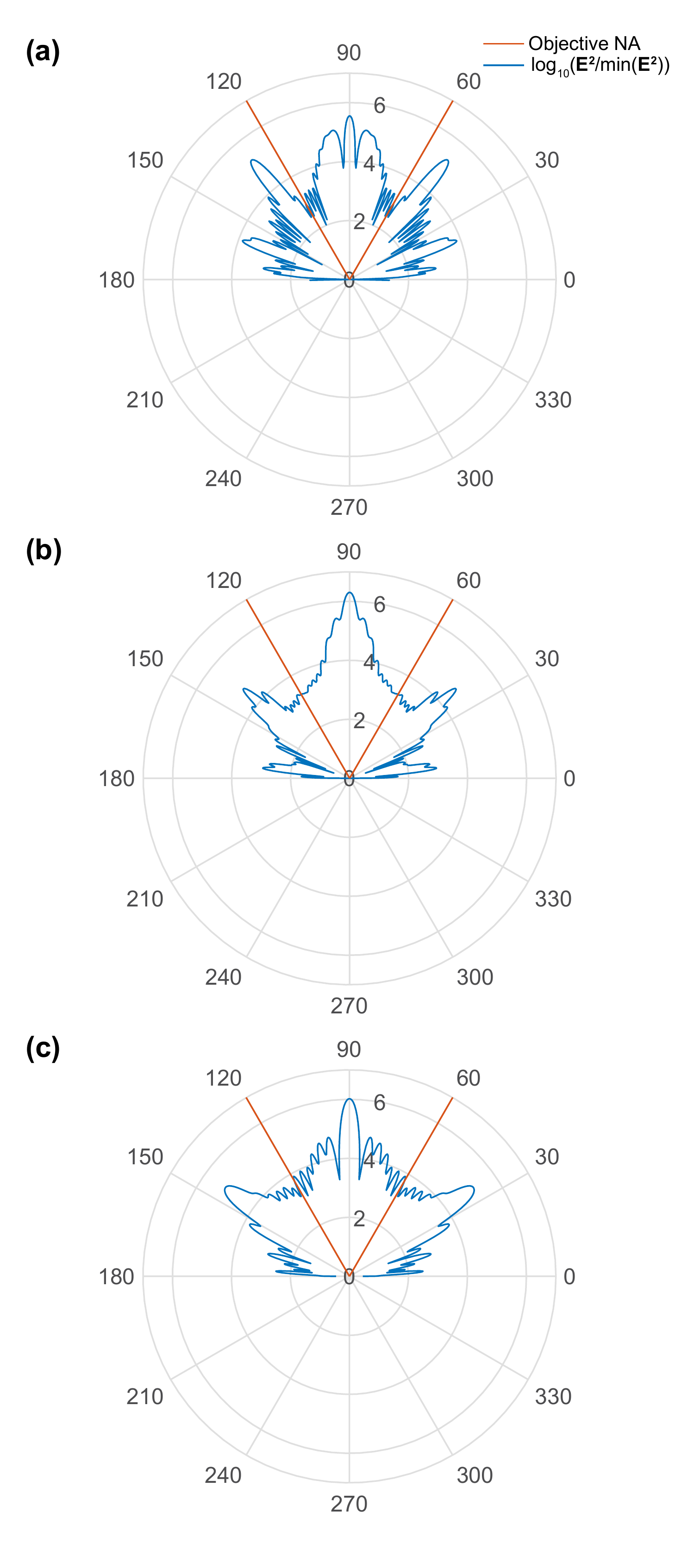}
    \caption{Far-field profiles (log scale) for the green filter normalized to the minimum field at 450nm wavelength. The wavelengths are  \textbf{(a)} 450 nm, \textbf{(b)} 550 nm, and \textbf{(c)} 650 nm. The acceptance angle of the objective is shown in orange, and all diffracted orders but zero fall beyond the objective's NA.} 
   \label{figs2}
 \end{center}
\end{figure}

As noted in the main text, the diffracted light from the filter is removed by angular or spatial
filtering. Figure \ref{figs2} shows the simulated far-field projections of a green filter alongside the NA of
the objective. In all cases the higher diffracted orders fall outside the NA of the objective

\subsection{Filter simulations}
The simulation model of the focusing filters was based on  single pixel filter with perfect matching layer (PML) in all boundaries (x = y = $-25 \mu m$ : $+25 \mu m$, z = $-4 \mu m$ : $+55 \mu m$) and total scattering field source (TSFS). The lower boundary of the source was placed $2\mu$m below the filter pixel inside the substrate. A frequency domain power monitor was placed in the focal plane of the filter. The focusing efficiency was calculated by normalizing poynting vector integration through a  circular area of$5 \mu m$ radius to the source power. The area was chosen with this size to match the experiments. Figures (\ref{figs4} \& \ref{figs5} ) show the incident angle effect on the optical power profile and focusing efficiency.

\section{SPECTRAL RESPONSE OF THE PHASE PLATE ALONE}
In addition to the complete pixel fabrication (phase-plate + pillars), We fabricated the phase-plate only and measured the focusing efficiency to study the effect of phase-plate on the filtering process. The results are shown in figure(\ref{figs5}).

\begin{figure}
\begin{center}
    \includegraphics[width=\textwidth]{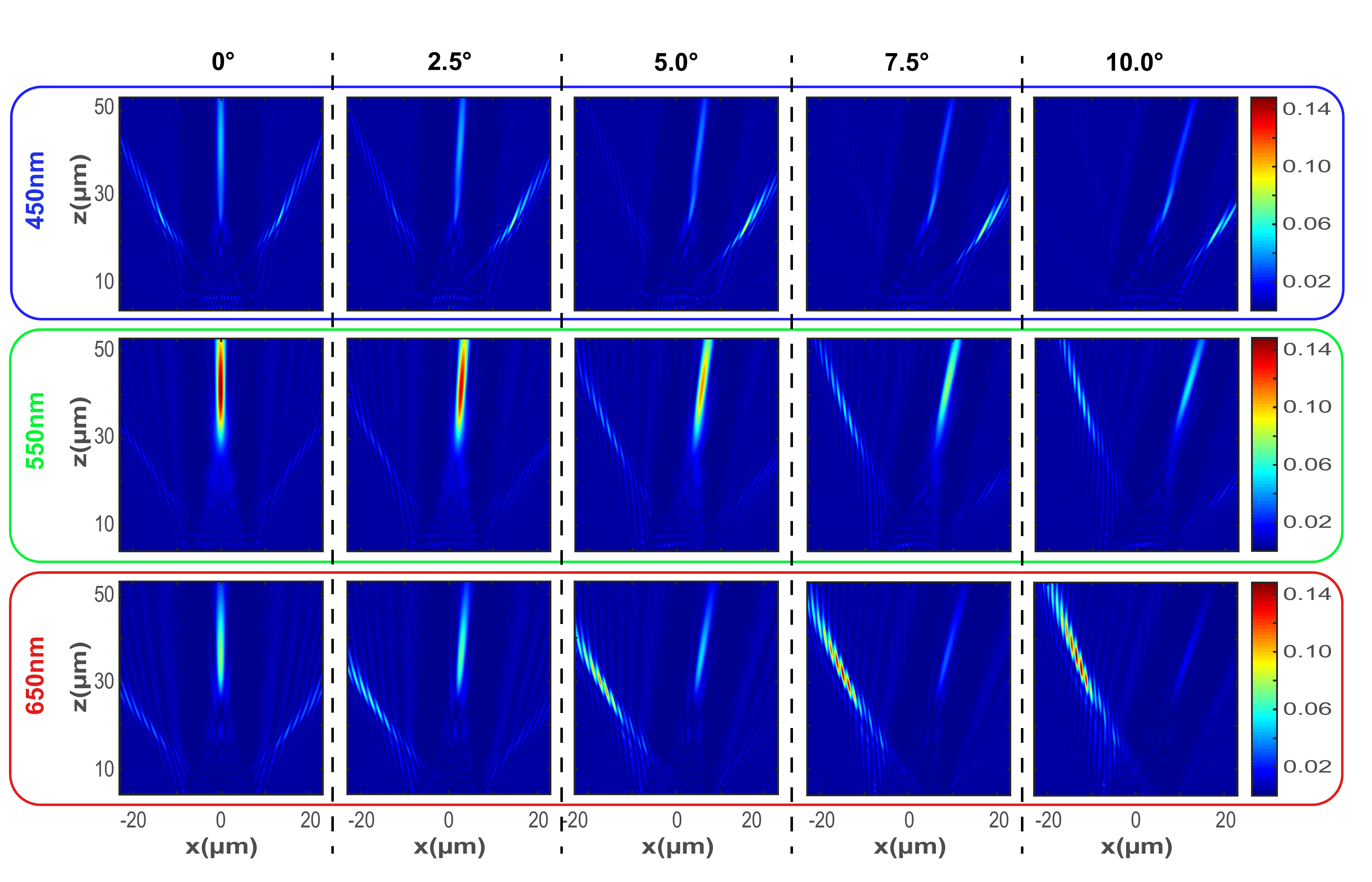}
    \caption{Optical power profile of a green filter in the xz-plane at different angles of incidence and different wavelengths as labeled on the figure. A change in the angle of incidence moves the focused spot as expected, and reduces the overall transmission, but has little effect on the relative transmitted power among the three wavelengths.}
   \label{figs3}
 \end{center}
\end{figure}

\begin{figure}
\begin{center}
    \includegraphics[width=\textwidth]{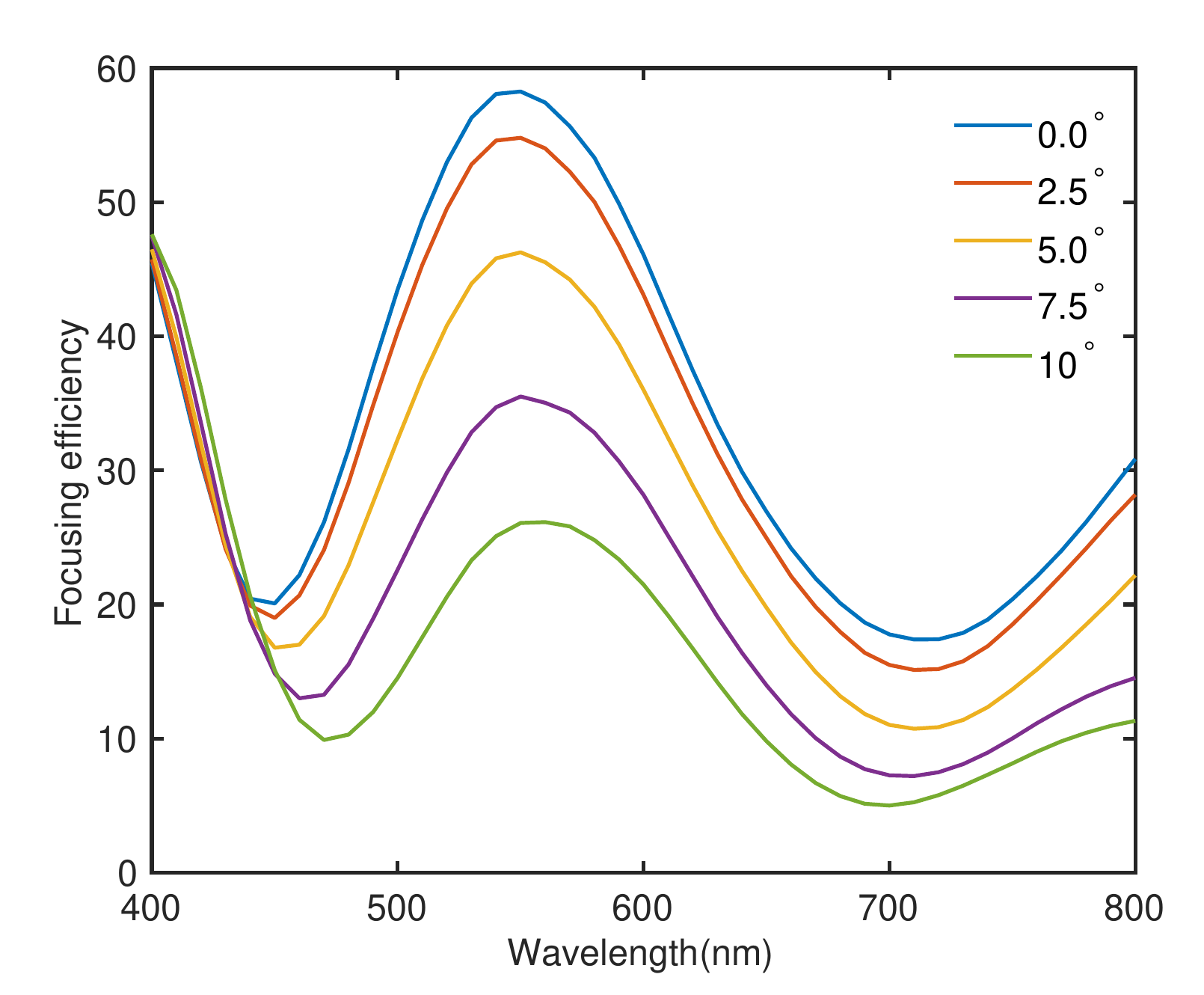}
    \caption{Simulated focusing efficiency of green filter at different angles of incidence. Although the overall transmission changes, the shift in the filter spectrum, and thus the color, is negligible}
   \label{figs4}
 \end{center}
\end{figure}
\begin{figure}
\begin{center}
    \includegraphics[width=\textwidth]{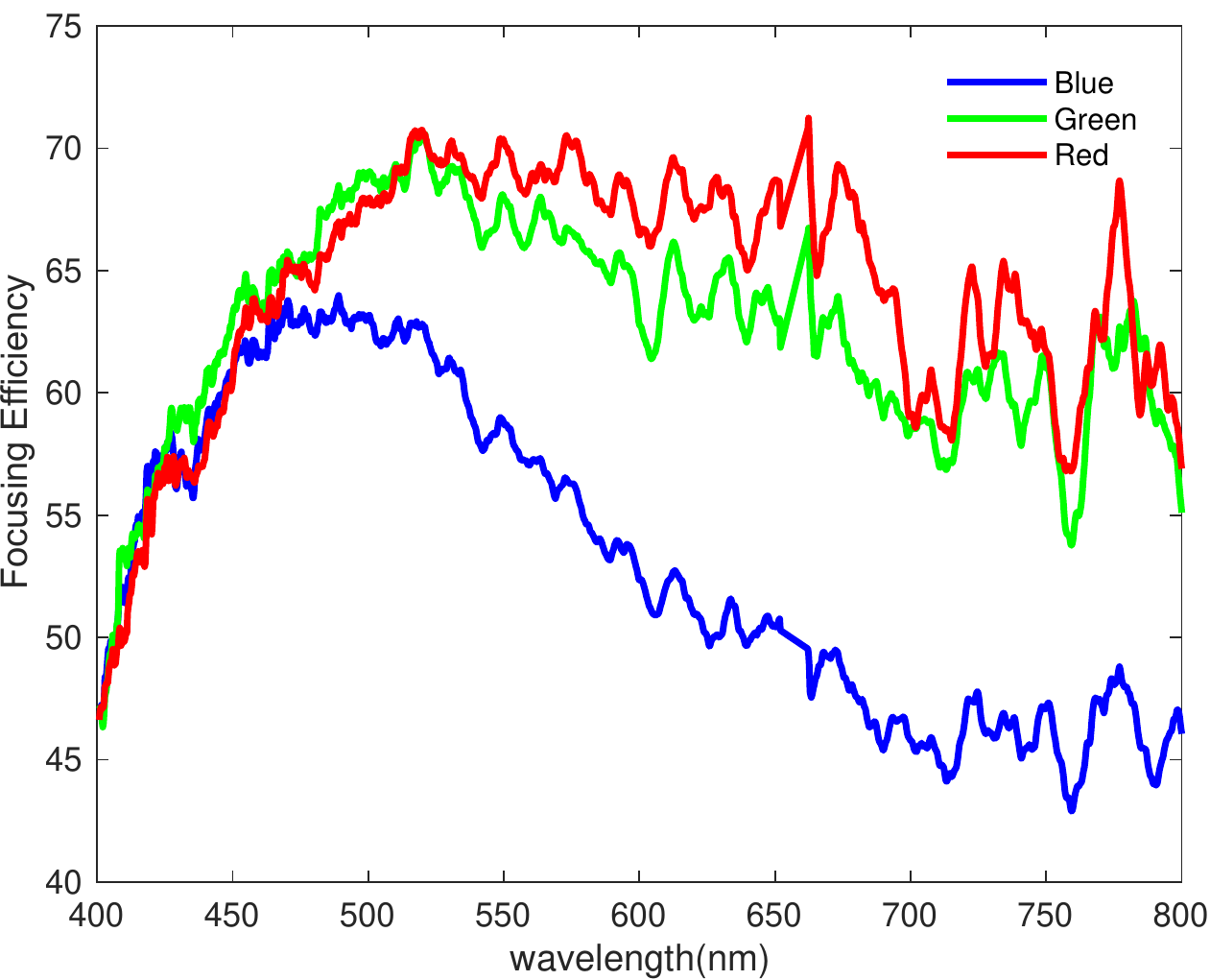}
    \caption{Focusing efficiency of phase-plate layer without pillars on top for the three main filters.  The phase plate has a relatively week effect on the green and red filters, but influences the blue filter significantly.}
   \label{figs5}
 \end{center}
\end{figure}

\end{document}